\documentclass[iop]{emulateapj}

\slugcomment{}

\shorttitle{A cosmic comb model of FRBs}
\shortauthors{Zhang}

\begin{document}


\title{A ``Cosmic Comb'' Model of Fast Radio Bursts}


\author{Bing Zhang}
\affil{Department of Physics and Astronomy, University of Nevada Las Vegas, NV 89154, USA}




\begin{abstract}
Recent observations of fast radio bursts (FRBs) indicate a perplexing, inconsistent picture. We propose a unified scenario to interpret diverse FRBs observed. A regular pulsar, otherwise unnoticeable at a cosmological distance, may produce a bright FRB if its magnetosphere is suddenly ``combed'' by a nearby, strong plasma stream towards the anti-stream direction. If the Earth is to the night side of the stream, the combed magnetic sheath would sweep across the direction of Earth and make a detectable FRB. The stream could be an AGN flare, a GRB or supernova blastwave, a tidal disruption event, or even a stellar flare. Since it is the energy {\em flux} received by the pulsar rather than the luminosity of the stream origin that defines the properties of the FRB, this model predicts a variety of counterparts of FRBs, including a possible connection between FRB 150418 and an AGN flare, a possible connection between FRB 131104 and a weak GRB, a steady radio nebula associated with the repeating FRB 121102, and probably no bright counterparts for some FRBs.
\end{abstract}





\section{Introduction}

The origin of fast radio bursts (FRBs) remains a mystery after one decade of observations \citep{lorimer07,thornton13,petroff15,champion16,masui15,keane16,spitler16,delaunay16,chatterjee17}. The distance scale of FRBs is finally settled to the cosmological range ($\sim$ Gpc) thanks to multi-wavelength follow-up observations that nail down the host galaxy of the repeating source that produced FRB 121102 \citep{chatterjee17,marcote17,tendulkar17}.

The collected data for FRBs so far are however perplexing and inconsistent. 1. Whereas the source of FRB 121102 clearly repeats \citep{spitler16,scholz16}, other FRBs so far have not been observed to repeat, despite dedicated follow-up observations for some of them \citep{petroff15b}; 
2. In the error circle of FRB 150418, a bright fading radio source was discovered in a galaxy at redshift $z=0.492$, which was claimed as the afterglow of the FRB \citep{keane16,zhang16b}. However, the radio source re-brightened later, leading to the suggestion that the radio source is a coincident background AGN not associated with the FRB \citep{williams16}. However, the chance probability to have such a highly variable AGN in the field of view of the FRB and to have a bright flare follow the FRB is low \citep{li16,johnston17}; 3. \cite{delaunay16} discovered a sub-threshold long GRB that was spatially and temporally associated with FRB 131104 \citep{ravi15}; 4. The repeating bursts from the FRB 121102 source are not associated with bright transient counterparts in other wavelengths, but the repeater is rather associated with a steady bright radio {\bf source} \citep{chatterjee17,marcote17,tendulkar17}.

Many FRB models have been proposed so far. However, no model can explain all the above observational facts. The catastrophic models \citep[e.g.][]{totani13,kashiyama13,falcke14,zhang14,zhang16a,wang16} cannot interpret the repeating FRB 121102. The previously favored super-giant-pulse model (in analogy to nanoshots from the Crab pulsar) \citep{cordes16,connor16} invokes nearby galaxy not at the cosmological distances and therefore are challenged by the fact that the repeater is located in a host galaxy at redshift $z= 0.193$ \citep{tendulkar17} (but see \citealt{katz16a,katz16b}). The magnetar giant flare model \citep{popov10,kulkarni14,katz16c} is in apparent conflict with the non-detection of bright radio pulse from the SGR 1806-20 giant flare \citep{tendulkar16}. Other repeating models \citep[e.g.][]{dai16,gu16} invoke specific physical conditions. These models also cannot account for the apparent association of FRB 131104 with the gamma-ray transient Swift J0644.5-5111 \citep{delaunay16}.  Models invoking more energetic magnetars as the source of FRBs have been proposed \citep{zhang14,murase16,piro16,murase17,gaozhang17,dai17,metzger17}. However, in order to account for both the repeater and the putative association of a GRB for FRB 131104 \citep{delaunay16}, both catastrophic and repeating mechanisms are needed. 

Here we propose an alternative mechanism to produce FRBs, which may give a unified interpretation of all FRBs observed so far. This model invokes an interaction between an astrophysical plasma stream and a foreground regular pulsar. The pulsar is otherwise non-detectable at the cosmological distance, but may produce an FRB as observed by an Earth observer when the plasma stream with significant ram pressure combs the magnetosphere towards the anti-stream direction and sweeps the direction of Earth.




\section{The model}

The condition for a cosmic comb is that the ram pressure of the stream is greater than the magnetic pressure of the pulsar magnetosphere. For a most general expression, we write the ram pressure in the relativistic form. The energy-momentum tensor of a perfect fluid is given by $T^{\mu\nu} = (\rho c^2 + e + p) u^\mu u^\nu + p g^{\mu\nu}$, where $g^{\mu\nu}$ is the metric tensor, $u^\mu=\gamma (1, \beta_x, \beta_y, \beta_z)$ is the dimensionless 4-velocity, $(\beta_x, \beta_y, \beta_z)$ is the dimensionless 3-velocity vector with amplitude $\beta=v/c$, $\gamma=(1-\beta^2)^{-1/2}$ is the Lorentz factor, $\rho$, $e$, and $p=(\hat\gamma-1) e$ are the mass density, internal energy density, and pressure of the fluid, respectively, $\hat\gamma$ is the adiabatic index, and $c$ is speed of light. For a fluid with co-moving mass density $\rho_0$, internal energy density $e$ and pressure $p$, the lab-frame energy density is $T^{00}=\gamma^2(\rho_0 c^2+e+p)-p=\gamma^2 \rho_0 c^2 +(\hat\gamma \gamma^2-\hat\gamma+1)e$. Subtracting the rest mass energy density one can obtain the ram pressure. The condition for a cosmic comb then reads
\begin{equation}
 (\gamma^2-1) \rho_0 c^2 + (\hat\gamma \gamma^2 - \hat\gamma+1) e > \frac{B^2}{8\pi}.
 \label{eq:cond}
\end{equation}
For a non-relativistic ($\gamma^2-1 \simeq \beta^2$), cold ($e \simeq 0$) flow, this is reduced to the familiar form of
\begin{equation}
 \rho v^2 > \frac{B^2}{8\pi}.
 \label{eq:cond-NR}
\end{equation}

The magnetosphere of a pulsar is enclosed within the light cylinder with radius $R_{\rm LC} = c/\Omega = cP/2\pi \sim (4.8 \times 10^9 {\rm cm}) P $, where $\Omega$ and $P$ are the angular frequency and period of the pulsar, respectively. The magnetosphere of a pulsar is significantly modified if the ram pressure exceeds the magnetic pressure at the light cylinder, $B_{\rm LC} \sim B_s (R_{\rm LC}/R)^{-3}$, where $R \sim 10^6$ cm is the radius of the neutron star, and a dipolar magnetic configuration has been assumed. The right hand side of Eqs.(\ref{eq:cond}) and (\ref{eq:cond-NR}) is therefore
\begin{equation}
 \frac{B_{\rm LC}^2}{8\pi} = \frac{B_s^2}{8\pi} \left(\frac{\Omega R}{c}\right)^{6} \simeq (3.4 \ {\rm erg \ cm^{-3}}) \ B_{s,12}^2 P^{-6},
 \label{eq:PB}
\end{equation}
where the median values of surface magnetic field $B_s = 10^{12} \ {\rm G}$ and period $P = 1$ s for Galactic radio pulsars have been adopted (throughout the paper, the convention $Q=10^n Q_n$ is adopted in cgs units). Noticing the strong dependence on $B_s$ and especially on $P$, one can conclude that cosmic combs more easily happen in slow and low-field pulsars.

For a relativistic blastwave such as a GRB afterglow or a relativistic AGN flare, since $e \gg \rho_0 c^2$ and $\hat\gamma \simeq 4/3$ due to relativistic shock heating, the left hand side of Eq.(\ref{eq:cond}) can be simplified as 
\begin{equation}
\frac{4}{3} \gamma^2 e \simeq \frac{4}{3} \gamma^3 n_{\rm ISM} m_p c^2 \simeq (2.0 \ {\rm erg \ cm^{-3}}) \ \gamma_1^3 n_{\rm ISM,0}
\label{eq:Pram1}
\end{equation} 
for a nominal value $\gamma = 10$ and $n_{\rm ISM} = 1 \ {\rm cm^{-3}}$, where $e = (\gamma-1) n_{\rm ISM} m_p c^2$. For a non-relativistic outflow such as a supernova explosion, a quasi-isotropic compact star merger ejecta, or a quasi-isotropic tidal disruption ejecta, the left hand side of Eq.(\ref{eq:cond-NR}) can be expressed as 
\begin{equation}
\rho v^2 = \frac{\Delta M v^2}{4\pi r^2 \Delta} =  (14.2 \ {\rm erg \ cm^{-3}}) \left(\frac{\Delta M}{M_\odot}\right) \beta_{-1}^2 r_{17}^{-2} \Delta_{16}^{-1}
\end{equation}
for an impulsive explosion (where $\Delta M$ is the ejecta mass, $\Delta$ is the thickness of the shell, $r$ is the distance from the center of explosion, and $v$ is the speed of the ejecta), or 
\begin{equation}
\rho v^2 = \frac{\dot M v}{4\pi r^2} =  (1.5 \ {\rm erg \ cm^{-3}}) \left(\frac{\dot M}{\rm M_\odot \ yr^{-1}}\right) \beta_{-1} r_{17}^{-2}
\end{equation}
for a continuous wind (where $\dot M$ is the wind mass loss rate). One can see that cosmic comb condition can be satisfied for a variety of systems if the pulsar is close enough to the source. For typical parameters, the ``horizon" of a cosmic comb is of the order of 0.1 pc. However, noticing the sensitive dependence of magnetic pressure on $P$ and $B$, the horizon can easily reach $\sim$ pc for slower and weaker-field pulsars.

If the cosmic comb condition is satisfied, when the plasma stream suddenly arrives, the pulsar magnetosphere is combed towards the anti-stream direction in a duration of $\sim R_{\rm sh} / v \sim (3  {\rm s}) \ R_{sh,10} \beta_{-1}$, where $R_{\rm sh} \sim R_{\rm LC}$ is the distance of the sheath from the pulsar (Fig.1, in analogy with the Earth magnetosphere). The strong ram pressure of the stream may trigger magnetic reconnections, which would quickly accelerate particles to relativistic speeds within a time scale much shorter than the combing time scale. The relativistic particles move along the magnetic field lines and flow out from the sheath to produce coherent radio emission. During the combing process, the sheath very quickly sweeps a near-$2\pi$ hemisphere solid angle. If the electrons (or electron-positron pairs) in the sheath move with a Lorentz factor $\gamma_e$, then the emission beam angle of the sheath would be $1/\gamma_e$, which is usually much smaller than the physical opening angle of the sheath. If the distance range over which emission into the characteristic radio band of the telescope (e.g. GHz range) is small, an observer at a random direction in the ``night'' side of the hemisphere would detect the signal in a duration
\begin{equation}
 \Delta t \sim \frac{R_{\rm sh}} {v \gamma_e} \simeq (3.3 \ {\rm ms}) R_{\rm sh,10} \beta_{-1}^{-1} \gamma_{e,3}^{-1}.
 \label{eq:Delta-t}
\end{equation}
This is consistent with the millisecond durations of the observed FRBs if $\gamma_e \gtrsim 10^3$ is satisfied.

The coherent radio emission mechanism is not specified. One interesting estimate is that with the nominal parameters, the curvature radiation frequency is
\begin{equation}
 \nu = \frac{3}{4\pi} \frac{c}{\rho} \gamma_e^3 \simeq (7.2\times 10^8 \ {\rm Hz}) \ \rho_{10}^{-1} \gamma_{e,3}^3,
\end{equation}
which falls into the range of FRB detection frequency (the curvature radius $\rho \sim R_{\rm sh}$ has been assumed). If the combing process may allow particles to emit in phase (the antenna mechanism), bright coherent radio emission can be generated. Alternatively, a cyclotron instability mechanism, as was proposed to interpret pulsar radio emission \citep{kazbegi91}, may operate near the light cylinder region to power the FRB. A sheath-origin coherent emission has been introduced to interpret radio emission of Pulsar B in the double pulsar system \citep{lyutikov04}. Here we assume that the combing process can produce bright coherent radio emission detectable from cosmological distances, and encourage a full investigation of the coherent mechanism of such a process in detail.

\begin{figure}
\plotone{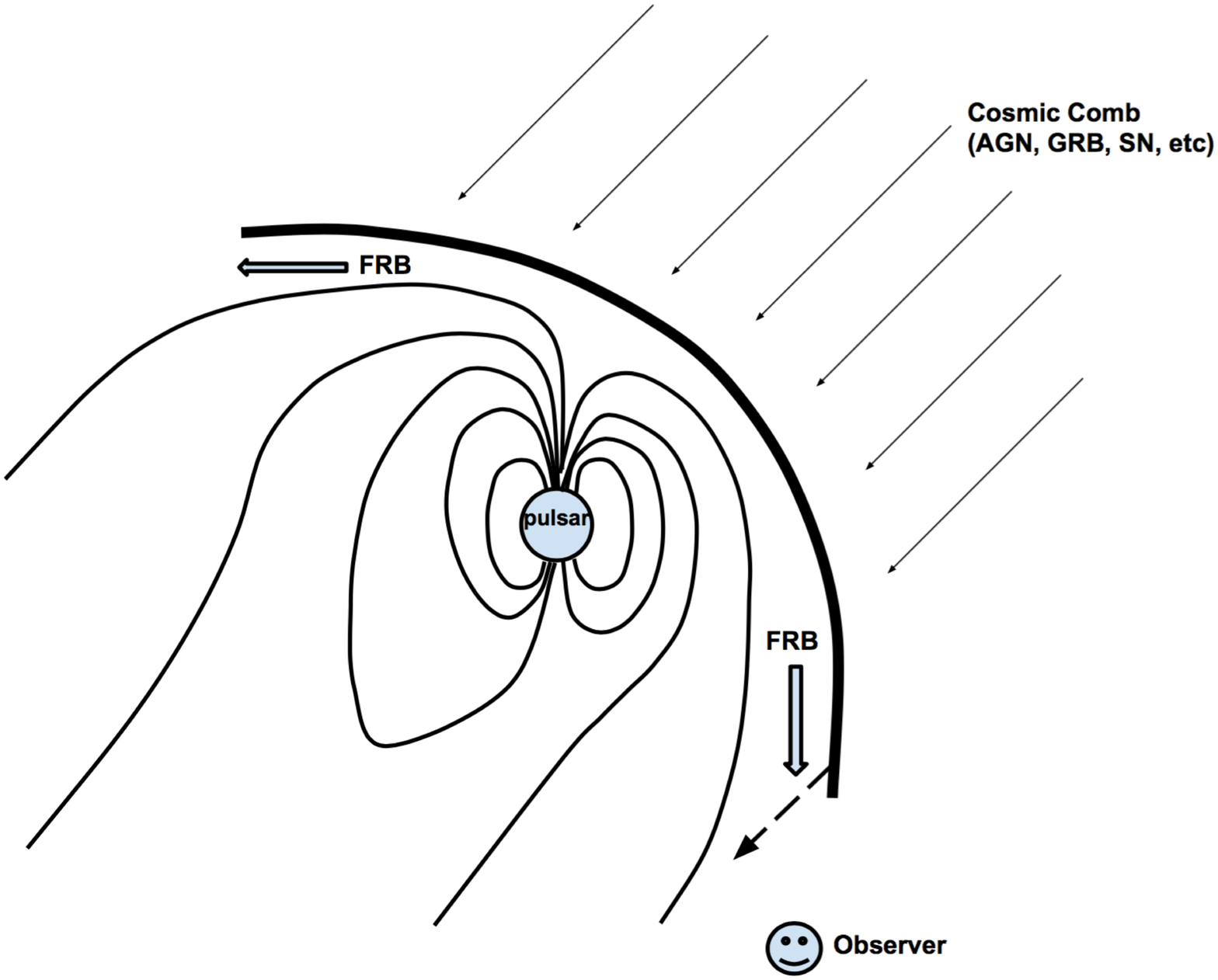}
\caption{A cartoon picture of a cosmic comb. An FRB is produced in the sheath region, which sweeps the line of sight during a short period of time defined by Eq.(\ref{eq:Delta-t}).}
\label{fig1}
\end{figure}

\section{Case studies}

This model can explain the puzzling properties of several FRBs in a unified way:

{\bf FRB 150418:} \cite{keane16} discovered a bright radio source starting from 2 hours after the FRB. The source dimmed in the next two observational epochs at 6 and 8 days after the FRB, respectively, and reached a quiescent level with a steady flux. \cite{williams16} discovered that the source re-brightened to the level of the original detection flux more than 300 days later, suggesting that the radio source is an AGN rather than the afterglow of the FRB. Long term monitoring of the source \citep{johnston17} suggested that the source is usually not in the high state. A Monte Carlo simulation suggests that the random probability of having the FRB to occur almost during the peak flux time of the AGN activity is very low,  i.e. $10^{-3}-10^{-4}$ \citep{li16}. Instead of attributing the AGN flare to an independent event from the FRB, we interpret FRB 150418 as emission from a combed pulsar by the AGN flare\footnote{Other mechanisms to connect an FRB with an AGN have been also suggested in the literature \citep[e.g.][]{romero16,zhang17}.}. A prediction is that FRB 150418 {\em may} repeat during another bright flare from the same AGN. However, not all flares may trigger additional FRBs from the same pulsar. This is because at the close distance ($<0.1$ pc) from the super-massive black hole, the pulsar must be undergoing orbital motion, so that there are occasions when the geometry does not work for the cosmic comb signal to be detectable from Earth. Within this picture, the galaxy at $z=0.492$ is indeed associated with FRB 150418, as is supported by the measured DM of the FRB \citep{keane16}.

{\bf FRB 131104:} \cite{delaunay16} discovered a sub-threshold, putative GRB that coincides with FRB 131104 both in spatial position and in time. A radio afterglow was not detected \citep{shannon17}, but the non-detection is consistent with the afterglow model if the ambient density is low (as expected from the NS-NS or NS-BH merger models) or the shock microphysics parameters are low \citep{murase17,gaozhang17,dai17}. The possible mechanisms to produce an FRB associated with a GRB include collapse of a supra-massive millisecond magnetar to a black hole \citep{zhang14}, which requires that the FRB appears near the end of an extended X-ray plateau; or a pre-merger electromagnetic processes \citep{zhang16a,zhang16b,wang16}, which requires that the FRB leads the burst. The latter scenario may be argued to marginally match the data \citep{dai17,gaozhang17}. However, there might be $\gamma$-ray emission already 7 seconds before the FRB according to the data. Furthermore, the Swift BAT was not pointing toward the source direction before $-7$ seconds with respect to the FRB \citep{delaunay16}. So it is likely that the FRB occurred during the process of a long-duration GRB. If so, known models are difficult to interpret the FRB. In the cosmic comb model, one requires that a pulsar is located at a distance $r > \gamma^2 c (7 \ {\rm s}) \sim 2 \times 10^{16} \ {\rm cm} \gamma_{2.5}^2$ away from the central engine in the direction of the jet (or at a closer distance if the line of sight is mis-aligned from the pulsar-engine direction). Considering a possible star forming region for a long GRB or a possible globular cluster for a NS-NS or NS-BH merger event, the chance probability to have a foreground pulsar from the GRB may not be small.

{\bf The repeater (FRB 121102):} The repeater is located in a star-forming dwarf galaxy at $z=0.193$ \citep{tendulkar17}. The source is associated with a radio source \citep{marcote17}, which is offset from the center of the galaxy \citep{tendulkar17}. A plausible scenario might be that the source of the FRB, likely a rapidly spinning magnetar, is at the center of the radio source and pumping energy to power a nebula \citep[e.g.][]{yang16,murase16,piro16,metzger17}. However, this model predicts an observable evolution of DM over the year time scale  \citep{piro16,metzger17,yang17}, which is marginally inconsistent with the non-detection of DM evolution of the repeating FRBs. Within the cosmic comb scenario proposed in this paper, the repeating bursts may originate from a foreground pulsar being episodically combed by an unsteady flow from a young supernova remnant. If the condition (\ref{eq:cond}) is marginally satisfied, the pulsar may relax to its normal magnetospheric configuration after a particular combing, but may be combed again and again when clumps with a higher ram pressure reach the pulsar magnetosphere region repeatedly. The pulsar is therefore observed to emit FRBs repeatedly. For a remnant with a finite width $\Delta$ and speed $v$, the repeating phase may last for $\Delta/v = 10^7 \ {\rm s} \Delta_{16} v_9^{-1}$. Since the repeater has been observed to repeat in a multi-year time scale, the remnant may be continuously energized by a central engine, likely a rapidly rotating neutron star. Since the FRB source is a foreground pulsar from the central source, the DM evolution could be much weaker depending on the geometry, consistent with the data. There is no direct observational evidence of ram pressure variation within a supernova remnant. However, for a nebular powered by continuous energy injection from a central engine (which is {\em not} the source of FRBs in the cosmic comb model), variation of ram pressure of the stream is expected. For a marginally satisfied comb condition envisaged here, a variation of ram pressure by a factor of a few would suffice to make a repeating FRB source as observed.

{\bf Other FRBs:} No counterparts have been claimed for other FRBs. Within the cosmic comb model, the ram pressure of the plasma stream essentially depends on the energy {\em flux} of the stream source. For example, a flare from a companion star (similar to a corona mass ejection event of the Sun) may provide a comparable ram pressure to a pulsar as the blastwave of a more distant GRB or supernova. As a result, one does not necessarily expect that all FRBs are associated with bright counterparts. It is possible that a fraction of FRBs may be associated with GRBs, supernovae, AGN flares, or tidal disruption events, while other FRBs may be combed by a stellar flare of a companion star so that no bright counterpart is detectable.

\section{Discussion}

We have proposed a scenario of FRBs that may interpret the perplexing, inconsistent FRB phenomenology within a unified framework. The magnetosphere of a slow, low-field pulsar may be combed by a cosmic plasma stream towards the direction of the Earth to make an FRB. The stream may originate from an explosion (a supernova, a GRB, or a neutron star merger event), an AGN flare, a tidal disruption event, or even a stellar flare from a binary companion of the pulsar.

The event rate density of FRBs is estimated as
$ \dot \rho_{\rm FRB} \simeq (5.7\times 10^3 \ {\rm Gpc^{-3} \ yr^{-1}}) ({D_z}/{3.4 \ {\rm Gpc}})^{-3} ({\dot N_{\rm FRB}}/{2500})$ ,
where $\dot N \sim 2500 \ {\rm day}^{-1}$ is the number of FRBs per day all sky with the current telescope sensitivity, and the typical value $D_z$ is scaled to $z=1$ \citep{zhang16a}. The cosmic comb model can easily account for such a large event rate density. This is because many explosions (e.g. supernovae and GRBs) and flares (from AGNs or companion stars) can potentially make FRBs if there are one or more pulsars in the vicinity of each plasma stream. In fact, the total event rate density of all the relevant explosions and flares exceed by orders of magnitude than the inferred FRB event rate density. A correct FRB event rate density may be obtained if the probability of having a slow, weak field pulsar in the vicinity is not negligibly small. In star forming regions (where GRBs and supernovae occur), globular clusters (where compact star mergers occur), and galactic center regions (where AGN flares occur), the stellar density can be as high as 100-1000 ${\rm pc^{-3}}$ \citep{genzel89,harris91,genzel10}. Since one old pulsar within $\sim$(0.1-several) pc from an explosion source or AGN would give a 50\% probability to make a cosmic comb towards Earth (as long as  Earth is in the night side hemisphere of the stream), one would expect a good fraction of explosions/AGN flares to trigger cosmic combs. Adding binary systems composed of an old pulsar and a flaring companion, the event rate would be even higher.\footnote{Within this picture, one may expect some of the Galactic binary systems would make FRBs. However, the event rate would be extremely low, i.e. at most $\sim$ 1 per century based on the observed FRB rate.} More detailed Monte Carlo simulations are needed to better estimate the event rate density of cosmic combs.

Supports to the cosmic comb model proposed in this paper may be obtained by a list of observational tests, e.g. a detection of repeating bursts from FRB 150418 and the localization of the bursting source to the spatially coincident AGN; more detections of FRBs associated with GRBs and afterglows; detections of FRBs that are associated with supernovae, tidal disruption events, and other cosmological transients.

On the other hand, it is entirely possible that multiple mechanisms are at play to make FRBs. Some other models have specific predictions that can be tested by future observations. For example, if future localizations of FRBs all reveal associations with bright radio nebulae, it would suggest a new-born millisecond magnetar as the source of FRBs \citep[e.g.][]{murase16,metzger17}, an FRB coincident with a gravitational wave (GW) signal due to a neutron star - neutron star (NS-NS), neutron star - black hole (NS-BH), or black hole - black hole (BH-BH) merger would suggest a direct connection between the FRB and the merger  \citep{totani13,zhang16a,wang16}\footnote{The charged BH merger model proposed by \cite{zhang16a} applies to all three types of mergers, since NSs are naturally charged; the mechanism of \cite{wang16} applies to mergers with at least one NS member; and the mechanism of \cite{totani13} applies to NS-NS mergers only. For NS-NS and NS-BH mergers, an FRB can be observed only in a small solid angle where FRB emission is not blocked by the dynamical ejecta launched during the merger process.}; and an FRB coincident with the disappearance of an extended plateau emission would suggest the ejection of a magnetosphere during the collapse of a supra-massive neutron star (``blitzar'') as the mechanism of FRBs \citep{zhang14,falcke14}.

\acknowledgments
I thank Sarah Burk-Spolaor, Zi-Gao Dai, Yuan-Pei Yang, and an anonymous referee for helpful comments.
This work is partially supported by NASA NNX15AK85G and NNX14AF85G.

\end{document}